\documentclass{article}
\usepackage{spconf,amsmath,graphicx,hyperref}

\usepackage{amssymb}
\usepackage{booktabs}
\usepackage{multirow} % for table multirow
\usepackage{soul} % pacakge for underline
\usepackage{silence}
\usepackage{xcolor}

\title{The Achilles' Heel of Angular Margins: \\A Chebyshev Polynomial Fix for Speaker Verification}

\name{
Yang Wang\textsuperscript{1}, Yiqi Liu\textsuperscript{1}, Chenghao Xiao\textsuperscript{2}, Chenghua Lin\textsuperscript{1}
}
\address{
\textsuperscript{1}The University of Manchester, UK, \textsuperscript{2}Durham University, UK
}
\begin{document}
%\ninept
%
\maketitle
\begin{abstract}
Angular margin losses, such as AAM-Softmax, have become the de facto in speaker and face verification. Their success hinges on directly manipulating the angle between features and class prototypes. However, this manipulation relies on the arccos function to recover the angle, introducing a significant yet overlooked source of training instability. 
% The derivative of arccos approaches infinity at its boundaries, leading to gradient peaks and optimisation difficulties, especially as the model converges and feature-prototype alignment becomes high. 
% Furthermore, this formulation struggles to generate a sufficiently sharp gradient for hard examples, where the feature-prototype alignment is moderately high. 
% In this work, we address this fundamental issue by proposing to replace the problematic arccos operation with its Chebyshev polynomial approximation. 
% This substitution not only yields an objective with a well-behaved gradient across the entire domain but also reshapes the loss landscape to apply a stronger corrective signal to hard-to-classify examples, leading to more effective optimisation. 
The derivative of arccos explodes at its boundaries, causing gradient peaks during optimisation. Furthermore, the formulation fails to generate a sufficiently sharp gradient for hard-to-classify examples. We address these issues by proposing ChebyAAM, a loss that replaces the arccos operation with its Chebyshev polynomial approximation. This substitution eliminates gradient explosion and applies a stronger corrective signal to hard examples, leading to more effective optimisation. 
% Extensive experiments on three benchmarks (VoxCeleb, SITW, and CN-Celeb) demonstrate that our approach effectively resolves the gradient instability, and leads to better performance. Our findings suggest that approximating angular operations, rather than calculating them explicitly, offers a more robust and effective path for designing future metric learning losses. 
% Our code will be publicly available.
Experiments on three benchmarks (VoxCeleb, SITW, and CN-Celeb) demonstrate that our method resolves the instability and consistently improves performance. Our work suggests that approximating angular operations, rather than calculating them explicitly, offers a more robust path for designing future metric learning losses. 
Code is available at \url{https://github.com/ExtraOrdinaryLab/vibe}.
\end{abstract}
\begin{keywords}
Polynomial approximation, margin-based loss, speaker verification
\end{keywords}

\section{Introduction}
\label{sec:intro}

Deep metric learning, particularly through angular margin losses like AAM-Softmax \cite{deng2019arcface}, has become the cornerstone of modern speaker verification \cite{chung20b_interspeech, wang23ha_interspeech, liu2024golden}. These methods excel by engineering a compact intra-class distribution and a separated inter-class distribution by enforcing a margin directly in the angular space. For instance, AAM-Softmax applies an additive margin $m$ to the angle $\theta$ between a feature and its class prototype, optimising for $\cos(\theta + m)$.

% Despite its success, this formulation conceals a critical flaw. To manipulate the angle, these methods must first recover it from the cosine similarity via the inverse cosine ($\arccos$) function. The instability originates here. The derivative of $\arccos(x)$ is $\frac{-1}{\sqrt{1 - x^2}}$, which approaches infinity as its input $x$ approaches $\pm 1$. As a model converges, the cosine similarity between an embedding and its target prototype naturally approaches 1. We empirically observed that the backward pass yields an exploding gradient, leading to numerical instability and hindering smooth optimisation. 
% Beyond this instability, a second, more subtle issue arises. As noted by the derivative surface in Figure~\ref{fig:loss_landscape}, the standard AAM-Softmax loss does not sufficiently sharpen the gradient for \textit{hard} examples (i.e., those with moderately high cosine similarities that are still far from the margin's target).
Despite its success, this formulation has two critical flaws rooted in its use of the inverse cosine ($\arccos$) function. First, the derivative of $\arccos(x)$, which is $\frac{-1}{\sqrt{1 - x^2}}$, explodes as the input $x \to 1$. As a model's embeddings align with class prototypes, their cosine similarity approaches 1, leading to exploding gradients and numerical instability. Second, the standard AAM-Softmax loss fails to sufficiently sharpen the gradient for \textit{hard} examples -- those with moderately high cosine similarities that are still far from the margin's target (see Figure~\ref{fig:loss_landscape}). This hinders effective optimisation on the most informative samples. 
To address both issues, we propose replacing the problematic $\arccos$ operation with its Chebyshev polynomial approximation. Our resulting loss function, ChebyAAM, retains the geometric principles of AAM-Softmax while creating a more stable loss landscape. By design, ChebyAAM eliminates the source of gradient explosion and provides a tailored gradient response that applies a higher penalty to hard examples, leading to more effective optimisation and state-of-the-art performance. Our contributions are two-fold: 
\begin{itemize}
    % \item We identify and analyse the gradient instability in $\arccos$-based angular margin losses.
    \item We identify and analyse the dual issues of gradient instability and insufficient penalisation of hard examples in $\arccos$-based angular margin losses.
    % \item We propose ChebyAAM, a stable and efficient loss function using polynomial approximation.
    \item We propose ChebyAAM, a stable and efficient loss function using polynomial approximation that resolves both problems.
    % \item We empirically demonstrate its superior stability and performance on three benchmarks (VoxCeleb, SITW, and CN-Celeb).
\end{itemize}

\section{Related Work}
\label{sec:related_work}

The evolution of speaker verification losses began with Softmax, which was insufficient for learning highly discriminative features for open-set tasks \cite{liu19f_interspeech}. A key advancement was the reformulation of the objective on a hypersphere by normalising features and weights \cite{wang2017normface}, making the loss dependent only on the angle $\theta$ between them. This paved the way for margin-based losses. A-Softmax \cite{liu2017sphereface} introduced a multiplicative angular margin ($m\theta$), while AM-Softmax \cite{wang2018cosface} used a simpler additive cosine margin ($\cos\theta - m$). AAM-Softmax \cite{deng2019arcface} proposed an additive angular margin ($\theta+m$), which has become a dominant method due to its geometric elegance and strong performance. 
However, the instability of AAM-Softmax has been implicitly acknowledged through heuristic solutions. Researchers have reported unstable convergence with large margins \cite{chung20b_interspeech} and divergence with lightweight models \cite{li2019airface, huang2020curricularface}. Common workarounds include margin annealing schedulers \cite{chen20253d} or pre-training with a standard Softmax loss. These methods treat the symptoms of instability. In contrast, our work identifies and resolves the root cause: the unbounded derivative of the $\arccos$ function.

\section{Methodology}
\label{sec:methodology}

\begin{figure*}[!t]
    \centering
    \includegraphics[width=\textwidth]{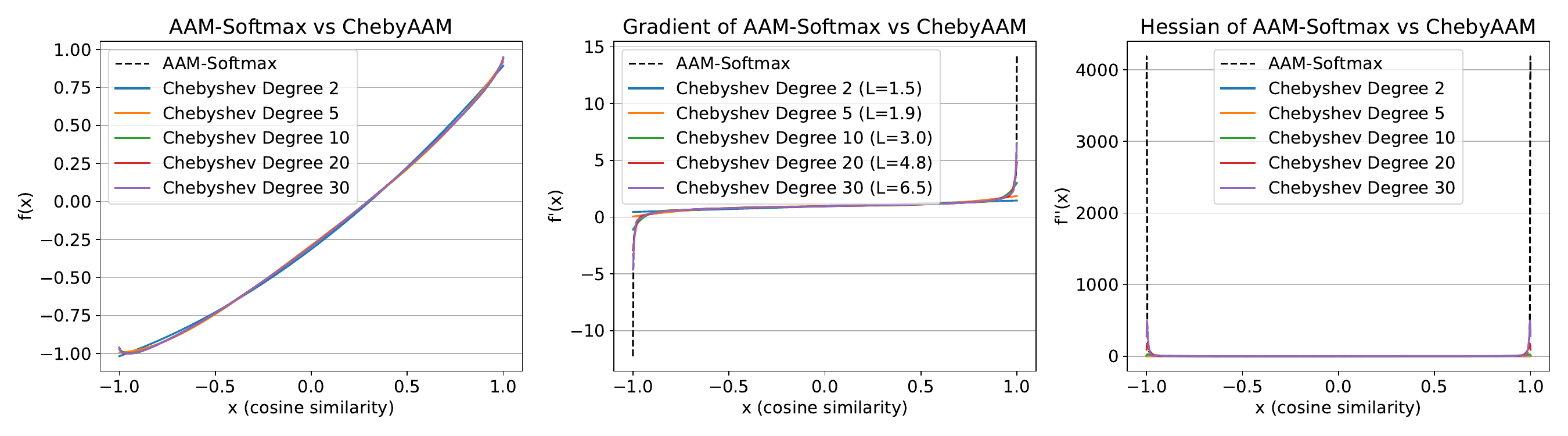}
    \caption{Function, gradient (first derivative), and Hessian (second derivative) plots for AAM-Softmax ($\psi$) and its Chebyshev approximations ($f_{\mathrm{cheb}}$) with margin $m=0.3$. The polynomial versions offer smoother derivatives and bounded Hessians, avoiding the pathological spikes seen in the exact function near $x=\pm 1$.}
    \label{fig:lipschitz_hessian}
\end{figure*}

\subsection{Motivation: The Instability of Arccos}
The core issue with angular margin losses like AAM-Softmax lies in the mandatory use of the $\arccos$ function to recover the angle $\theta$ before applying a margin. The derivative of $\arccos(x)$, given by $\frac{d}{dx}\arccos(x) = - \frac{1}{\sqrt{1 - x^2}}$, becomes unbounded as $x \to \pm 1$. During training, as embeddings become well-aligned with their class prototypes, the cosine similarity $x$ approaches 1, causing the gradient to explode. This is the primary source of the observed training instability.

\subsection{Proposed Polynomial Approximation}
\label{sec:polynomial_approximation}

To resolve this, we replace the composite function $\psi(x, m) = \cos(\arccos(x) + m)$ in AAM-Softmax with its Chebyshev polynomial approximation, $f_{\mathrm{cheb}}(x, m)$. Chebyshev polynomials are ideal for this as they minimise the maximum approximation error over the interval $[-1, 1]$. The expansion is given by:
\begin{equation}\label{eq:chebyshev}
    f_{\mathrm{cheb}}(x, m) = \frac{1}{2} a_0 + \sum_{k \ge 1} a_k T_k (x)
\end{equation}
where the coefficients are

\begin{equation}\label{eq:coefficients}
\begin{cases} 
a_0 = - \frac{2\sin(m)}{\pi}, a_1 = \cos(m) \\
% a_1 = \cos(m) \\
a_{2k+1} = 0 & \forall \quad k \ge 1 \\
a_{2k} = \frac{2\sin(m)}{\pi} \bigl( \frac{1}{2k-1} - \frac{1}{2k+1} \bigl) & \forall \quad k \ge 1
\end{cases}
\end{equation}
and the Chebyshev polynomials of the first kind $T_k$ are defined by 

\begin{equation}
    T_k(x) = \cos(k \arccos(x)) \quad \text{for} \quad x \in (-1, 1)
\end{equation}

Beyond stability, ChebyAAM is also efficient. The evaluation of the polynomial is performed using Clenshaw's algorithm \cite{clenshaw1955note}, a numerically stable and computationally inexpensive recurrence that requires only multiply-add operations. 
% \textcolor{blue}{After checking the later part, can we measure the actual compute time for both or is it just theoretical? Will be much stronger if it's actually fast.} 
Intuitively, the polynomial approximation form typically has simpler, well-behaved derivatives compared to the combination of cosine and $\arccos$. 
This can lead to more stable gradient computations (middle plot in Fig.~\ref{fig:lipschitz_hessian}), reducing issues like vanishing/exploding gradients. 
More stable training can potentially lead to a better final model. 
Additionally, a polynomial approximation can make it easier to analyse properties like the Lipschitz constant of the loss function (i.e. supremum of the norm of the gradient): 

\begin{equation}
    L_{\mathrm{lipschitz}} = \max_{x \in [-1, 1]} \| \nabla_x f_{\mathrm{cheb}}(x, m) \|
\end{equation}
or the curvature (i.e., Hessian properties). 
For example, in Fig.~\ref{fig:lipschitz_hessian} right plot, we can see spikes in the Hessian near branch cuts \cite{chong_teaching_2021} at $x = \pm 1$. This is tied to the derivative of $\arccos$. 

\subsection{Gradient of Chebyshev-based AAM-Softmax}

Given Chebyshev expansion (Eq.~\ref{eq:chebyshev} and Eq.~\ref{eq:coefficients}), the differentiation of the series, $\nabla_x f_{\mathrm{cheb}}(x, m)$, can be derived:

\begin{equation}\label{eq:first_derivative}
    \cos(m) + \sum_{k \ge 1} \frac{4k \sin(m)}{\pi} \bigl( \frac{1}{2k-1} - \frac{1}{2k+1} \bigl) U_{2k-1}(x)
\end{equation}
where $U_k(x)$ is the Chebyshev polynomial of the second kind, defined by: 

\begin{equation}
    U_k(x) = \frac{\sin((k+1) \arccos(x))}{\sqrt{1 - x^2}} \quad \text{for} \quad x \in (-1, 1)
\end{equation}

The Hessian of the Chebyshev series, $\nabla_x^2 f_{\mathrm{cheb}}(x, m)$, is further obtained, written as:

\begin{equation}\label{eq:second_derivative}
    - \sum_{k \ge 1} a_k k \frac{k T_k(x) \sqrt{1 - x^2} - x \sin(k\arccos(x))}{(1 - x^2)^{3/2}}
\end{equation}

By approximating the angular margin in AAM-Softmax with a Chebyshev series (Eq.~\ref{eq:chebyshev}), we obtain closed-form expressions for the gradient (Eq.~\ref{eq:first_derivative}) and Hessian (Eq.~\ref{eq:second_derivative}). 
These representations reveal that both the gradient and Hessian are composed of weighted sums of well-understood Chebyshev basis functions $T_k(x)$ and $U_k(x)$. 
Due to the rapid decay\footnote{In default, we use a polynomial of degree 30, and the first few coefficients $a_k$ are calculated: $a_0 \approx -0.1265$, $a_1 \approx 0.98007$, $a_2 \approx 0.08433$, $a_3 \approx 0$, $a_4 \approx 0.01687$.} of the Chebyshev coefficients for smooth functions, the local behaviour of the loss (in terms of both sensitivity and curvature) is dominated by a few low-order terms. 
This structure not only allows us to derive uniform error bounds on the approximation but also provides insights into the optimisation dynamics, such as ensuring favourable curvature properties in regions critical for convergence. 
Such analytic control over the loss's derivatives is instrumental in both designing and theoretically justifying robust training algorithms in deep speaker recognition.

\begin{figure*}[tp]
    \centering
    \includegraphics[width=1.0\linewidth]{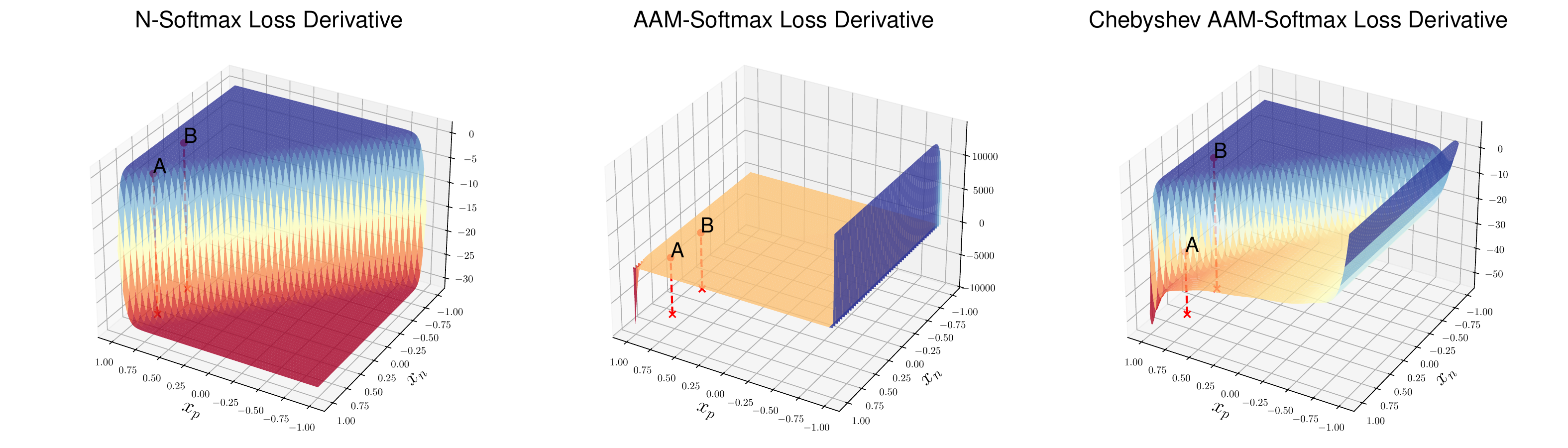}
    \caption{Comparison of the derivative surfaces of N-Softmax (left), AAM-Softmax (middle), and Chebyshev-based AAM-Softmax (right) with respect to the true-class logit. The ``hard'' example (point A, where $s_p \approx s_n$) and the ``easy'' example (point B, where $s_p \ll s_n$) end up with similar magnitudes of gradients for the N-Softmax and AAM-Softmax losses, but differ more for the Chebyshev approach. Chebyshev AAM-Softmax can show a bigger derivative gap between A and B. It effectively can have a steeper slope near certain values, thus providing a higher penalty for ``hard'' examples relative to ``easier'' ones.}
    \label{fig:loss_landscape}
\end{figure*}

To understand the benefits of our approach, we visualise the loss gradients in a binary classification scenario with positive ($s_p$) and negative ($s_n$) logits (Fig~\ref{fig:loss_landscape}). A ``hard'' example occurs when the logit difference is small (e.g., Point A: (0.8, 0.8)), while an ``easy'' example has a large separation (e.g., Point B: (0.8, 0.2)).

Intuitively, hard examples should receive a much larger gradient to facilitate learning. However, for both N-Softmax and AAM-Softmax, the gradient magnitudes at points A and B are not dramatically different. 
% \textcolor{blue}{I think this part serves as a better motivation which will be good to also highlight introduction and the narrative in general, as opposed only the gradient exposion thing. What do you think?} 
This is because the derivative of the standard cross-entropy loss saturates, and the angular margin of AAM-Softmax does not sufficiently sharpen the gradient for moderately high cosine similarities.

In contrast, ChebyAAM exhibits a more pronounced difference in gradient magnitude between hard and easy examples. The polynomial approximation reshapes the loss landscape, creating different curvature in critical regions. This allows ChebyAAM to generate a steeper gradient for hard examples, providing a stronger corrective signal where it is most needed and leading to more effective optimisation. This tailored gradient response explains the improved performance observed in our experiments.

\section{Experimental Setup}
\label{sec:setup}

\noindent\textbf{Datasets.}~~We trained our primary model on the VoxCeleb2 development set and tested on the VoxCeleb1-O, VoxCeleb1-E, and VoxCeleb1-H trial lists \cite{nagrani2017voxceleb, chung2018voxceleb2}. To assess out-of-domain generalisation, we also evaluated on the SITW core-core and core-multi lists \cite{mclaren2016speakers}. For Mandarin performance, a separate model was trained on CN-Celeb1/2 and tested on the CN-Celeb1 evaluation list \cite{fan2020cn, li2022cn}.

\noindent\textbf{Model and Training.}~~We used the ECAPA-TDNN \cite{desplanques20_interspeech}, CAM++ \cite{wang23ha_interspeech}, SimAMResNet \cite{qin2022simple} architectures. All models were trained for 70 epochs using an SGD optimiser with a warmup cosine learning rate scheduler (peak learning rate 0.2) and a batch size of 256. Input features were 80-dimensional Fbanks from 3-second audio chunks. We applied data augmentation including speed perturbation and the addition of noise\footnote{\url{https://www.openslr.org/17}} and reverberation\footnote{\url{https://www.openslr.org/28}}.

\noindent\textbf{Evaluation.}~~Following \cite{chen20253d}, we report Equal Error Rate (EER) and Minimum Detection Cost Function (mDCF) with $P_{\mathrm{target}}=0.01$ using cosine similarity scoring, without any back-end score normalisation, to provide a clear assessment of the loss function's performance.

\begin{table*}[!t]\centering
\resizebox{\textwidth}{!}{ % use this if the table is too large
\begin{tabular}{ccccccccccccc}\toprule
\multirow{2}{*}{\textbf{Objective}} & \multicolumn{2}{c}{\textbf{VoxCeleb-O}} & \multicolumn{2}{c}{\textbf{VoxCeleb-E}} & \multicolumn{2}{c}{\textbf{VoxCeleb-H}} & \multicolumn{2}{c}{\textbf{SITW-core-core}} & \multicolumn{2}{c}{\textbf{SITW-core-multi}} & \multicolumn{2}{c}{\textbf{CN-Celeb1-Eval}} \\\cmidrule(lr){2-13}
& EER\% & mDCF & EER\% & mDCF & EER\% & mDCF & EER\% & mDCF & EER\% & mDCF & EER\% & mDCF \\
\midrule
A-Softmax & 1.1222 & 0.1286 & 1.4290 & 0.1750 & 2.8732 & 0.2954 & 2.0230 & 0.1880 & 5.5397 & 0.3021 & 8.3526 & 0.5054 \\
AM-Softmax & 0.8453 & 0.1068 & 1.0872 & 0.1301 & 2.1448 & 0.2131 & 1.6676 & 0.1477 & 5.0174 & 0.2542 & 7.6641 & 0.4489 \\
AAM-Softmax & 0.8988 & 0.1010 & 1.0066 & 0.1132 & 1.9681 & 0.1989 & 1.6402 & 0.1280 & 4.6083 & 0.2332 & 8.0181 & 0.4526 \\
ChebyAAM (ours) & \textbf{0.8453} & \textbf{0.0837} & \textbf{0.9793} & \textbf{0.1127} & \textbf{1.9008} & \textbf{0.1960} & \textbf{1.3122} & \textbf{0.1262} & \textbf{4.3459} & \textbf{0.2247} & \textbf{7.1099} & \textbf{0.4048} \\
\bottomrule
\end{tabular}
}
\caption{A comparative study between ChebyAAM and its baselines using ECAPA-TDNN. The best performance is in \textbf{bold}.}\label{tab:main}
\end{table*}

\begin{table}[!t]\centering
\resizebox{\linewidth}{!}{ % use this if the table is too large
\begin{tabular}{lrrrrr}\toprule
\textbf{Model} &\textbf{Loss} &\textbf{Vox1-O} &\textbf{Vox1-E} &\textbf{Vox1-H} \\\midrule
\multirow{2}{*}{ECAPA-TDNN} &AAM-Softmax &0.8988 &1.0066 &1.9681 \\
&ChebyAAM &\textbf{0.8453} &\textbf{0.9793} &\textbf{1.9008} \\
\midrule
\multirow{2}{*}{SimAMResNet34} &AAM-Softmax &\textbf{0.6808} &0.8554 &1.6419 \\
&ChebyAAM &0.6861 &\textbf{0.8533} &\textbf{1.5888} \\
\midrule
\multirow{2}{*}{CAM++} &AAM-Softmax &0.9145 &1.0993 &2.1366 \\
&ChebyAAM &\textbf{0.7709} &\textbf{0.9838} &\textbf{1.9466} \\
\bottomrule
\end{tabular}
}
\caption{Comparison of EER (\%) on VoxCeleb1 test sets for AAM-Softmax and ChebyAAM across different backbones. 
% ChebyAAM consistently improves or matches performance, demonstrating its broad applicability and model-agnostic benefits.
}\label{tab:architectures}
\end{table}

\begin{table}[!t]\centering
\resizebox{\linewidth}{!}{ % use this if the table is too large
\begin{tabular}{ccccccc}\toprule
\multirow{2}{*}{\textbf{Degree}} & \multicolumn{2}{c}{\textbf{Vox1-O}} & \multicolumn{2}{c}{\textbf{Vox1-E}} & \multicolumn{2}{c}{\textbf{Vox1-H}} \\\cmidrule(lr){2-7}
& EER\% & mDCF & EER\% & mDCF & EER\% & mDCF \\
\midrule
2 & 0.8666 & 0.1029 & 0.9710 & 0.1123 & 1.9115 & 0.1921 \\
5 & 0.8031 & 0.1098 & 1.0144 & 0.1146 & 1.9626 & 0.2070 \\
10 & 0.7818 & 0.1026 & 0.9952 & 0.1132 & 1.9400 & 0.1925 \\
20 & 0.8932 & 0.1234 & 0.9872 & 0.1192 & 1.9553 & 0.1985 \\
30 & 0.8453 & 0.0837 & 0.9793 & 0.1127 & 1.9008 & 0.1960 \\
40 & 0.8956 & 0.0912 & 1.0167 & 0.1284 & 1.9127 & 0.1977 \\
50 & 0.8933 & 0.0964 & 1.0199 & 0.1256 & 1.9102 & 0.1974 \\
\bottomrule
\end{tabular}
}
\caption{Ablation study on the impact of the Chebyshev polynomial degree on verification performance across the VoxCeleb1 test sets using ECAPA-TDNN. 
% Higher degrees provide a more faithful approximation of the true angular margin function, as detailed in \S\ref{sec:polynomial_approximation}.
}\label{tab:degree}
\end{table}

\section{Results and Discussion}

\subsection{Main Results on VoxCeleb, SITW, and CN-Celeb}

All experiments in this section were conducted with a margin of 0.3. The choice of 0.3 was the result of a grid search over the range [0.1, 0.5]. As detailed in Table~\ref{tab:main}, ChebyAAM consistently outperforms all baseline margin-based losses across the full range of evaluation datasets, with particularly notable gains on the more challenging SITW datasets.

% This performance gain stems from enhanced training stability. We demonstrate this explicitly in Table~\ref{tab:stability} by testing both losses under more aggressive conditions: a high margin ($m=0.5$) and a \textit{monotone squashing} transformation. This transformation applies a power function with an exponent of 1.5 to the positive-class logit to artificially accelerate its convergence towards the unstable boundary of 1. As shown, the standard AAM-Softmax baseline failed under both settings, leading to NaN losses. In contrast, ChebyAAM remained perfectly stable, proving its robustness against the gradient explosion that plagues $\arccos$-based methods. This stability allows the model to learn a more effective and generalisable embedding space.

\subsection{Generalisability Across Architectures}

To demonstrate that the benefits of ChebyAAM are not specific to a single backbone, we evaluated on three different model architectures, shown in Table~\ref{tab:architectures}. ChebyAAM consistently improves or matches the performance of the AAM-Softmax across all models. For the ECAPA-TDNN and CAM++, ChebyAAM provides a clear reduction in EER. On the SimAMResNet34, it delivers on-par results, with a slight improvement on the Vox1-H set. This demonstrates that replacing the $\arccos$ operation is a fundamentally beneficial and model-agnostic strategy for improving training.

% \textcolor{blue}{any results on ablating Chebyshev degree? Agree will be good}

\subsection{Ablation on Polynomial Degree}

We conducted an ablation study, presented in Table~\ref{tab:degree}, to analyse the impact of the Chebyshev polynomial's degree on performance. The results show that even a low-degree polynomial provides a strong baseline that is competitive with the standard AAM-Softmax. As the degree increases, the polynomial provides a more faithful approximation of the original angular margin function, which generally leads to improved verification performance. In our experiments, a degree of 30 yielded the best overall results, justifying its use in our main experiments. As the degree increases further to 50, the performance begins to degrade and converge back towards that of the original AAM-Softmax. This suggests that higher degree approximation may start to replicate the gradients of the original $\arccos$ function, diminishing the regularisation benefit of the lower-degree polynomials.

\section{Conclusion}
\label{sec:conclusion}

% In this paper, we identified a fundamental source of instability in popular angular margin losses: the unbounded derivative of the $\arccos$ function. We proposed ChebyAAM, a novel loss function that replaces the problematic trigonometric operation with a stable Chebyshev polynomial approximation. Our experiments on VoxCeleb, SITW, CN-Celeb benchmarks demonstrate that ChebyAAM not only resolves the gradient instability but also leads to superior verification performance, especially on in-the-wild data. This work suggests that approximating, rather than explicitly calculating, angular operations is a more robust path for designing future metric learning losses.

In this paper, we identified two key weaknesses in popular angular margin losses stemming from the $\arccos$ function: gradient instability due to its unbounded derivative, and an ineffective gradient response for hard-to-classify examples. We proposed ChebyAAM, which resolves these issues by replacing the problematic trigonometric operation with a stable Chebyshev polynomial approximation. Our experiments across VoxCeleb, SITW, and CN-Celeb benchmarks demonstrate that ChebyAAM eliminates training instability and reshapes the loss landscape, leading to more effective optimisation and superior speaker verification performance. 
% This work suggests that approximating, rather than explicitly calculating, angular operations is a more robust and effective path for designing future metric learning losses.

\section{Acknowledgment and Disclosure}

Yang Wang gratefully acknowledges the support provided by Automated Analytics. We disclose that LLM was used only for polishing and proofreading. The authors reviewed all suggestions to ensure accuracy and that no figures or experimental results are LLM-generated. This use complies with the ICASSP policy on acceptable use of LLMs.

% To start a new column (but not a new page) and help balance the last-page
% column length use \vfill\pagebreak.
% -------------------------------------------------------------------------
\vfill
\pagebreak

% References should be produced using the bibtex program from suitable
% BiBTeX files (here: strings, refs, manuals). The IEEEbib.bst bibliography
% style file from IEEE produces unsorted bibliography list.
% -------------------------------------------------------------------------
\bibliographystyle{IEEEbib}
\bibliography{strings,refs}

\end{document}